\documentclass[aps,pra,twocolumn,superscriptaddress]{revtex4-1}

\usepackage[utf8]{inputenc} 
\usepackage[T1]{fontenc} 
\usepackage[english]{babel}
\usepackage[autostyle=true]{csquotes} 
\usepackage{xcolor}
\usepackage{graphicx}
\usepackage{array}
\usepackage{amsmath} 
\usepackage{amssymb} 
\usepackage{amsfonts} 
\usepackage{amsbsy}
\usepackage{mathrsfs} 
\usepackage{dsfont} 
\usepackage{dcolumn} 
\usepackage{upgreek} 
\usepackage{bm} 
\usepackage{bbm}
\usepackage{placeins} 
\usepackage{physics} 
\usepackage[colorlinks,linkcolor=blue,citecolor=blue,urlcolor=blue]{hyperref}
\usepackage[most]{tcolorbox}
\usepackage{subfigure}
\usepackage{float}
\usepackage[sort&compress]{natbib}
\usepackage[dvips]{epsfig}
\usepackage{hhline}
\usepackage{multirow}
\usepackage[normalem]{ulem}
\newcommand{\amend}[1]{\textcolor{purple}{#1}}
\newcommand{\remove}[1]{\textcolor{orange}{\xout{(#1)}}}

\newcommand{\dfpt}{DFPT}

\def\ai	        {{\em ab--initio}}

\renewcommand{\[}{\left[}
\renewcommand{\]}{\right]}
\renewcommand{\(}{\left(}
\renewcommand{\)}{\right)}

\def\dg         {\dagger}
\def\rar        {\rightarrow}

\newcommand{\eq}[1]{\begin{align}#1\end{align}}
\newcommand{\ml}[1]{\begin{multline}#1\end{multline}}

\newcommand{\seq}[1]{\begin{subequations}#1\end{subequations}}
\newcommand{\sst}[2]{\substack{#1\\#2}}
\newcommand{\seql}[2]{\begin{subequations}\label{#1}#2\end{subequations}}
\newcommand{\mll}[2]{\begin{multline}\label{#1}#2\end{multline}}
\newcommand{\eql}[2]{\begin{align}\label{#1}#2\end{align}}
\newcommand{\eqgl}[2]{\seq{\label{#1}\begin{gather}#2\end{gather}}}
\newcommand{\stkout}[1]{\ifmmode\text{\sout{\ensuremath{#1}}}\else\sout{#1}\fi}
\newcommand{\average}[1]{\left\langle #1 \right\rangle}

\newcommand{\bps}[1]{\biggl[ #1 \biggr]}
\newcommand{\bpr}[1]{\biggl( #1 \biggr)}

\newcommand{\lab}[1]{\label{#1}}

\newcommand{\ul}[1]{\underline{#1}}

\newcommand{\e}[1]{Eq.~\eqref{#1}}
\newcommand{\es}[2]{Eqs.~\eqref{#1}--\eqref{#2}}
\newcommand{\elab}[2]{Eq.(\ref{#1}#2)}

\newcommand{\fig}[1]{Fig.\ref{#1}}

\newcommand{\evalat}[2]{\left.#1\right|_{#2}}

\newcommand{\app}[1]{Appendix\,\ref{#1}}
\newcommand{\h}[1]{\hat{#1}}
\newcommand{\ti}[1]{\tilde{#1}}
\newcommand{\wh}[1]{\widehat{#1}}

\newcommand{\ocite}[1]{Ref.\cite{#1}}

\def\Re{{\rm Re}}
\def\Im{{\rm Im}}
\newcommand{\p}{\prime}           
\def\ga         {\alpha}
\def\gb         {\beta}
\def\gc         {\gamma}
\def\gC         {\Gamma}
\def\gd         {\delta}
\def\gD         {\Delta}
\def\gee        {\epsilon}
\def\gl         {\lambda}

\def\go         {\omega}
\def\gO         {\Omega}
\def\gr         {\rho}

\def\gp         {\phi}

\def\kk		{{\mathbf k}}
\def\qq		{{\mathbf q}}

\newcommand{\di}{\mathrm{d}}
\newcommand{\im}{\mathrm{i}}
\def\callD{\mbox{$\mathcal{D}$}}

\def\callL{\mbox{$\mathcal{L}$}}

\def\callP{\mbox{$\mathcal{P}$}}
\def\callQ{\mbox{$\mathcal{Q}$}}

\def\callT{\mbox{$\mathcal{T}$}}

\newcommand{\cnrism} {Istituto di Struttura della Materia and Division of Ultrafast Processes in Materials (FLASHit) of the National Research Council, via Salaria Km 29.3, I-00016 Monterotondo Stazione, Italy}

\def\smII{\app{sec:SM_H}}

\def\smIII{\app{sec:SM_EOMs}}

\def\smIV{\app{sec:g_dressing}}

\def\smV{\app{sec:laplace_vs_fourier}}

\def\smVI{\app{sec:regularization}}

\def\smVII{\app{sec:QPH}}

\newcommand{\mysec}[1]{\section{#1}}
\newcommand{\fddfpt}{~\cite{Lazzeri2006,Saitta2008,Calandra2010}}
\renewcommand{\amend}[1]{#1}
\renewcommand{\remove}[1]{}

\begin{document}

\title{
Non--adiabatic effects lead to the breakdown of the semi--classical phonon picture 
}
\author{Andrea Marini}
\affiliation{\cnrism}
\begin{abstract}
Phonon properties of realistic materials are routinely calculated within the Density Functional Perturbation Theory\,(DFPT). This is a semi--classical
approach where the atoms are assumed to oscillate along classical trajectories immersed in the electronic Kohn--Sham system. 
In this work I demonstrate that, in metals, non--adiabatic effects induce a deviation of the 
DFTP phonon frequencies from the quantistic solution of the Dyson equation.
A deviation that increases with the phonon energy width reflecting the breakdown of the semi--classical DFPT description. The final message is that non--adiabatic
phonon effects can be described only by using a fully quantistic approach.
\end{abstract}
\date{\today}
\maketitle

\mysec{Introduction}
\lab{sec:intro}
The research on the physics induced or mediated by the lattice vibrations is crucial in many and disparate fields of modern solid--state physics: infrared
spectroscopy, Raman, neutron-diffraction spectra, thermal transport are just a few of them~\cite{Stefano2001}.  Density Functional
Theory\,(DFT)~\cite{R.M.Dreizler1990} and Density Functional Perturbation Theory(\dfpt)~\cite{Stefano2001,Gonze1995,Gonze1997a} have emerged as successful and
widely used approaches to calculate the structural, electronic properties and also atomic dynamics in a fully \ai, framework. DFT and \dfpt\, are, nowadays,
available in many public scientific codes\cite{Giannozzi2017,Gonze2009} and routinely used to calculate phonon frequencies and related properties. 

Within DFT and \dfpt\, the atoms are treated classically and  the atoms are assumed to move around the minimum of the Born--Oppenheimer
surface~\cite{Tavernelli2015}, with the electrons tightly bound to the atoms during their oscillations.  In metallic materials the electronic excitations
resonant with the phonon frequency cause a retardation between the electronic and atomic oscillations. This retardation is a non--adiabatic effect that
induces, for example, the broadening of the phonon energy.

Indeed, from the experimental point of view it is well--known that the phonon peaks observed in the inelastic X--Ray scattering~\cite{Shukla2003} or in the Raman
spectra~\cite{Ferrante2018} have an intrinsic energy width.  This energy width is a 
natural concept within the Many--Body Perturbation Theory\,(MBPT)~\cite{Leeuwen2013,Marini2023,Giustino2017,Leeuwen2004a,Marini2015,Stefanucci2023,Marini2018},
while the actual possibility to describe it in a semi--classical theory like \dfpt\, is still debated.
Some works\fddfpt have used perturbation theory to propose a frequency dependent extension of \dfpt\,(FD--DFPT). This
theory 
leads to the introduction of a frequency dependent and non--hermitian dynamical matrix whose eigenvalues, the phonons frequencies, are complex. 
FD--DFPT provides, thus, a picture conceptually equivalent to MBPT where non--adiabatic effects induce a renormalization and broadeding of the
phonon energies. Moreover
the enormous simplicity of FD--DFPT compared to the more involved MBPT approaches and its availability in many \ai\, codes, has favored the application of the
semi--classical \dfpt\, phonon concept beyond the adiabatic regime. The FD--DFPT has been applied to calculate, for examples: phonon
widths~\cite{Shukla2003,Lazzeri2005,Calandra2010}, dynamical Kohn anomalies\cite{Piscanec2004a,Lazzeri2006}, Raman
spectra~\cite{Saitta2008,Ferrante2018,Pisana2007/03//print,Caudal2007} and non--adiabatic Born effective charges~\cite{PhysRevLett.128.095901,Binci_2021}.

These works have cemented the idea that a description based on a semi--classical representation of the atomic degrees of freedom is formally equivalent to the more
involved  Many--Body approach, with the difference that FD--DFPT relies on the change of the electronic density while MBPT requires to solve complicated
equations written in terms of non--local phonon and electron Green's functions. 

In this work I mathematically demonstrate that, within non--adiabatic Density Functional Perturbation Theory,
phonon widths are strictly zero.  This implies that
 semi--classical atomic oscillations never decay in time, even when they are resonant with quantistic electron--hole pairs. I show
that an infinite phonon lifetime is necessary to ensure that the dynamics conserves the total energy.  The FD--DFPT approach will be demonstrated
to correspond to the energy violating, and thus nonphysical, solution of the \amend{time--dependent\,(TD)} DFPT equation.  I conclude the work by comparing the TD--DFPT and MBPT
phonon frequencies. I show that the two solutions diverge as the MBPT phonon width increases thus demonstrating that non--adiabatic effects can be
described only by using a fully quantistic approach.

\mysec{Semi--classical trajectories and quantum fluctuations}
\label{sec:H_and_EOM}
In order to describe non--adiabatic effects I rewrite DFPT in the time domain. I start from the perturbed DFT Hamiltonian written in second quantization:
\ml{
\h{H}=\sum_{i} \gee_{i} \h{\gr}_{ii}+ 
\sum_{\gl}\bps{\frac{\go_{\gl}}{2}\h{p}_\gl^\dag \h{p}_\gl+\frac{\(\go_\gl-C^{e-n}_\gl\)}{2}\h{u}^\dag_\gl\h{u}_\gl\\+
\sum_{ij} g_{ji}^{\gl} \Delta\h{\gr}_{ij} \h{u}_{\gl}}+\sum_{ij} \Delta V^{Hxc}_{ji} \h{\gr}_{ij}.
\lab{eq:H.1}
}
This Hamiltonian has been introduced by several authors\cite{Marini2023,Giustino2017,Leeuwen2004a,Marini2015,Stefanucci2023,Marini2018} to discuss the connection
between DFTP and MBPT. The procedure to derive \e{eq:H.1} from the full DFT Hamiltonian is outlined in the \smII.

\e{eq:H.1} is written in terms of the single--particle DFT electronic and DFPT phononic energies, $\gee_i$ and $\go_\gl$. $\h{\gr}_{ij}$ is the density matrix
and $\h{u}_\gl$ and $\h{p}_\gl$ are  the atomic displacement and momentum. $g^\gl_{ji}$  is the bare electron--phonon potential that appears together with  the
variation of the the Hartree plus exchange--correlation potential\,(Hxc), $\Delta V^{Hxc}_{ji}$.  $\h{\gr}_{ij}$ is written in terms of  the electronic creation
and annihilation operators: $\h{c}^\dag_i/\h{c}_j$ and  $\Delta \h{\gr}_{ij}=\h{c}^\dag_i\h{c}_j-\average{ \h{c}^\dag_i\h{c}_j}$.

In \e{eq:H.1} $C^{e-n}_\gl$ is the electron--nuclei\,(e--n) component of the reference DFPT phonon dynamical matrix. As it has been explained in
\ocite{Marini2023,Marini2015,Stefanucci2023} this terms is already included in the $\go_\gl$ definition and needs to be removed in order to avoid double counting
effects.

The Hamiltonian $\h{H}$ induces a time--dependent dynamics of all operators, electronic and atomic. These equations are derived in details in the \smIII.

In the case of the atomic displacement operator we have
\mll{eq:EOM.1}
{
\h{\callD}_\gl\(t\)\h{u}_{\gl}\(t\)=i\sum_{ij} g_{ji}^\gl \int_{0}^t e^{i\Delta\gee_{ij}\(t-\tau\)} \\
\[\ul{\h{\gr}}\(\tau\),\ul{g}^\gl\h{u}_\gl\(\tau\)+\ul{\Delta V}^{Hxc}\[\gr\]\]_{ij}
}
with $\h{\callD}_\gl\(t\)=\frac{1}{\go_\gl}\bpr{\frac{\di^2}{\di t^2}-\go_\gl\(C^{e-n}_\gl-\go_\gl\)}$. $t=0$ is the initial time, underlined quantities are
matrices in the single--particle basis, and $\Delta\gee_{ij}=\gee_i-\gee_j$.

The role of the $\ul{\Delta V}^{Hxc}$ potential is to dress the e--p interaction, as demonstrated in the \smIV.  In practice this means that we can remove
$\ul{\Delta V}^{Hxc}$ from \e{eq:EOM.1} and replace, in the commutator, $\ul{g}^\gl$ with the {\em screened} and frequency dependent electron--phonon interaction, 
$\ti{\ul{g}}^\gl\(\go\)=\ul{\ul{\gee}}^{-1}\(\go\)\ul{g}^\gl$, with $\ul{\ul{\gee}}^{-1}$ the dielectric tensor. For simplicity here we ignore the time dependence of
$\ul{\ul{\gee}}^{-1}$ and assume $\ti{\ul{g}}^\gl\(\go\)\approx \ti{\ul{g}}^\gl\(\go=0\)$\cite{LK}.

If we now take the average of both sides of \e{eq:EOM.1} we note that it appears the term $\average{\ul{\h{\gr}}\(t\)\h{u}_\gl\(t\)}$ \amend{appears on the
right--hand side\,(RHS)}. This
term prevents \e{eq:EOM.1} to be written in terms of the electronic density matrix $\ul{\gr}\(t\)=\average{\ul{\h{\gr}}\(t\)}$ and atomic trajectory
$u_\gl\(t\)=\average{\h{u}_\gl\(t\)}$ because
\eql{eq:EOM.2}
{
 \average{\ul{\h{\gr}}\(t\)\h{u}_\gl\(t\)}=\ul{\gr}\(t\)u_\gl\(t\)+\average{\ul{\Delta\h{\gr}}\(t\)\Delta\h{u}_\gl\(t\)}.
} 
In \e{eq:EOM.2} $\Delta{\h{u}}\(t\)=\h{u}\(t\)-u\(t\)$. The first term in \e{eq:EOM.2}
corresponds to the classical mean--field approximation and the second term represents the quantum corrections.  From a physical point of view \e{eq:EOM.2} makes
clear that the mean--field term describes the trajectory\,($\average{\h{u}_\gl\(t\)}$) while the second term describes the
fluctuations\,($\average{\Delta\ul{\h{\gr}}\(t\)\Delta\h{u}_\gl\(t\)}$) around the classical trajectories.  The term
$\average{\Delta\ul{\h{\gr}}\(t\)\Delta\h{u}_\gl\(t\)}$ can be written in terms of the phonon propagator and
self--energy\cite{Marini2023,Giustino2017,Leeuwen2004a,Marini2015,Stefanucci2023,Marini2018}.

The semi--classical picture corresponds to assume $\average{\ul{\h{\gr}}\(t\)\h{u}_\gl\(t\)}\approx \ul{\gr}\(t\)u_\gl\(t\)$.  It is {\em semi}--classical
because the atomic dynamics is described by a trajectory ($u_\gl\(t\)$) while the electronic sub--system is treated fully quantistic.  

As we are in the harmonic approximation we can further  assume that the atomic displacements are tiny enough to approximate  $\gr_{ij}\(t\)\sim f_i\gd_{ij}$,
with $f_i$ the electronic occupations. It follows that
\eqgl{eq:EOM.3}
{
\h{\callD}_\gl\(t\)u_{\gl}\(t\)=-\int_{0}^t C_\gl\(t-\tau\)u_\gl\(\tau\)\di\tau,\\
C_\gl\(t\)=-\im \sum_{ij}\Delta f_{ij}g_{ji}^\gl \ti{g}^\gl_{ij} e^{\im\Delta\gee_{ij}t},
}
with $\Delta f_{ij}=f_i-f_j$. \e{eq:EOM.3} is the Time--Dependent DFPT equation of motion whose solution is defined in terms of the the boundary conditions at $t=0$. Here we define
$u_\gl\(0\)=u^0_\gl$ and $\frac{\di}{\di t}u_\gl\(0\)=v^0_\gl$.

\mysec{Adiabatic and time--dependent Density--Functional Perturbation Theory}
\label{sec:DFPT_and_TDDFPT}
The adiabatic DFPT can now be obtained from \e{eq:EOM.3} by assuming $\Delta\gee_{ij}\gg \go_\gl$. This implies that during the rapid electronic oscillations the atoms do not
move and, \amend{on the RHS} of \e{eq:EOM.3}, $u_\gl\(\tau\) \approx u_\gl\(t\)$.  In this way  the \amend{RHS} of \e{eq:EOM.3} can be calculated analytically to give
\eql{eq:DFPT.1}
{
 \h{\callD}_\gl\(t\)u_{\gl}\(t\)=\[- C^{st}_\gl +\sum_{ij}\frac{\Delta f_{ij}g_{ji}^\gl \ti{g}^\gl_{ij}}{\Delta\gee_{ij}}
e^{\im\Delta\gee_{ij}t}\]
u_{\gl}\(t\),
}
with $C^{st}_\gl=\sum_{ij}\frac{\Delta f_{ij}g_{ji}^\gl \ti{g}^\gl_{ij}}{\Delta\gee_{ij}}$. As $\Delta\gee_{ij}\gg \go_\gl$ \e{eq:DFPT.1} can be
time integrated over an electronic period much shorter than the phonon period, where the integral of the last term \amend{on the RHS} of \e{eq:DFPT.1} vanishes. This
means that, within the adiabatic approximation, $\go_\gl$ represents the frequency of the slow oscillation of the solution of \e{eq:DFPT.1} when $C^{e-n}_\gl=C^{st}_\gl$ confirming that DFPT corresponds to the
take the static\,(adiabatic) limit of the TD--DFPT kernel.

\begin{figure}[t!]
{\centering
\includegraphics[width=\columnwidth]{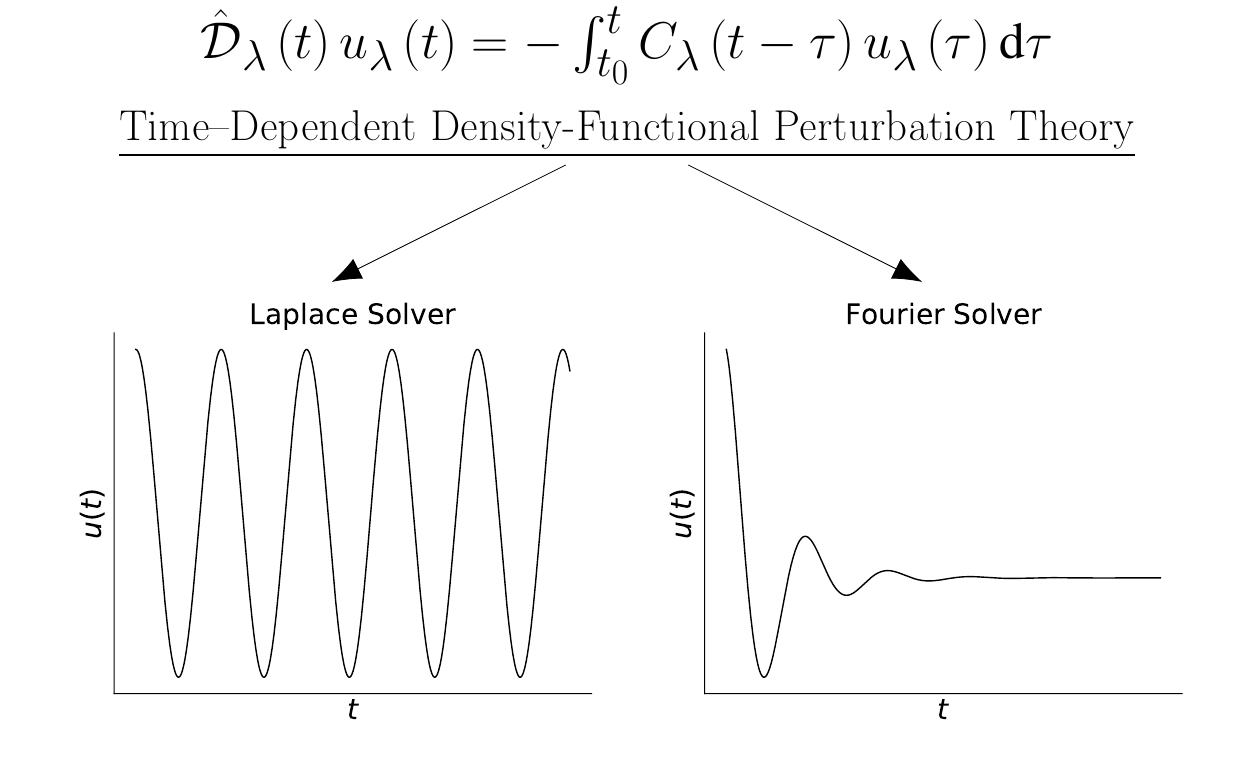}
}
\caption{
The two solvers (Laplace and Fourier) of the TD--DFPT equation, \e{eq:EOM.3}, are schematically compared.  $u\left(t\right)$ is the atomic displacement
function.  While the Laplace approach leads to persistent oscillations the Fourier solution produces an unphysical damping that violates the energy
conservation.  The Fourier solution corresponds to the frequency--dependent DFPT proposed by \ocite{Lazzeri2006,Saitta2008,Calandra2010}
\remove{and currently implemented in many \ai\, DFPT codes publicly released}.
}
\label{fig:1}
\end{figure}
In metals the adiabatic ansatz breaks down as the metallic electronic transition and atomic oscillation energies are similar, $\Delta\gee_{ij}\sim \go_\gl$.
Therefore non--adiabatic effects appear and, in \e{eq:EOM.3}, it is not possible to assume $u_\gl\(\tau\)\sim u_\gl\(t\)$.

The TD--DFPT phonon frequencies correspond, therefore, to the solution of \e{eq:EOM.3}. This is a second--order Volterra linear integro--differential
equation~\cite{Wazwaz2011} with a separable kernel\,($e^{\im\Delta\gee_{ij}\(t-\tau\)}$). The Volterra equations are the subject of an intense mathematical
research activity as they appear in many physical contexts like, for example, the dynamics of viscoelastic materials~\cite{AlabauBoussouira2008} or applications
of physical engineering~\cite{Zak2016}. 

A crucial property of \e{eq:EOM.3} is that, being based on the mean--field approximation, leads to an energy conserving dynamics. This means that if $E=\average{\h{H}}$
the solution of \e{eq:EOM.3} must lead to a time independent and constant $E$. 

To solve \e{eq:EOM.3} I use the Laplace transformation, which is commonly used to solve the free quantum harmonic oscillator equation~\cite{Wazwaz2011,boas_mathematical_2006}
\eql{eq:TDDFPT.1}
{
 f\(z\)=\callL[f]\(z\)=\int_0^{\infty} e^{-z t} f\(t\) d\,t.
}
When $z=-\im \go$ \e{eq:TDDFPT.1} reduces to the Fourier transformation. We consider here both transformations because, as it will
demonstrated shortly, the use of the Fourier will lead to the FD--DFPT theory proposed in \fddfpt, and widely used in the literature,
which predicts the non--adiabatic effects to induce a finite phonon energy indetermination, corresponding to complex phonon frequencies.

We now proceed to apply \e{eq:TDDFPT.1} to \e{eq:EOM.3} and we note that,  when $z=-\im \go$ (Fourier), $\callL[\h{\callD}_\gl u_\gl]\(-\im \go\)$ is well
defined only and only if $u_\gl\(t=\infty\)=0$.  The mathematical reason is that, in order to
transform the time differential operator, we need to use the integration by parts method which requires the integrand of \e{eq:TDDFPT.1} to vanish at
$t=\infty$ (\smV). This means, in practice, that the Fourier case can be used only by applying a tiny exponential prefactor, $e^{-\eta t}$, and send $\eta$ to zero at
the end of derivation. 
We thus consider $f\(z\)$ and $f\(-\im \go +\eta\)$, with $\eta\rar 0$.  

By using \e{eq:TDDFPT.1} it follows that
\seql{eq:TDDFPT.2}
{
\eq
{
 u_\gl\(z\)=\frac{z u_\gl^0 +v_\gl^0}{z^2+\go^2_\gl+\go_\gl \gD C_\gl\(z\)},
}
with $\gD C_\gl\(z\)=C_\gl\(z\)-C_\gl^{e-n}$ and
\eq
{
 C_\gl\(z\)=-\im \sum_{ij}g_{ji}^\gl \ti{g}_{ij}^\gl \frac{\Delta f_{ij}}{z-\im\Delta\gee_{ij}}.
}
}
$u_\gl\(t\)$ can be analytically obtained from $u_\gl\(z\)$ by using
the Bromwhich integral~\cite{boas_mathematical_2006} method that involves a complex plane integral of $u_\gl\(z\)$. From \e{eq:TDDFPT.2} it follows that the solution of
\e{eq:EOM.3} is equivalent to find the poles of $u_\gl\(z\)$. In the Laplace case this corresponds to find the zeros of\
\seql{eq:TDDFPT.3}
{
\eq{
 z^2   +\go^2_\gl+\go_\gl \gD C_\gl\(z\)=0,
}
while in the Fourier case we have
\eq{
 -\go^2+\go^2_\gl+\go_\gl \lim_{\eta\rar 0} \gD C_\gl\(-\im\go+\eta\)=0.
}}
We now remind that  $\lim_{\eta\rar 0}  \frac{1}{\go+\Delta\gee_{ij}+\im\eta}=\callP\[\frac{1}{\go+\Delta\gee_{ij}}\]-\im\pi\gd\(\go+\Delta\gee_{ij}\)$,
with $\callP$ the Cauchy principal and real value. Therefore it follows that
\eql{eq:TDDFPT.4}
{
 \lim_{\eta\rar 0} \gD C_\gl\(-\im\go+\eta\)=\callP\[\gD C_\gl\(-\im\go\)\]+\im \gC_\gl\(\go\),
}
with
\eql{eq:TDDFPT.5}
{
 \gC_\gl\(\go\)=-\pi \sum_{ij}g_{ji}^\gl \ti{g}_{ij}^\gl \Delta f_{ij}\gd\(\go+\Delta\gee_{ij}\).
}
\e{eq:TDDFPT.4} implies that the poles defined by \elab{eq:TDDFPT.3}{b} are complex. In the Laplace case, instead, \amend{as demonstrated in \smVI,}
\remove{we observe that by, definition,} $C_\gl^{e-n}\in\mathbb R$ and $\gD C_\gl\(z\)=\callP\[C_\gl\(z\)\]$.
It follows that the Laplace solution leads to imaginary poles (In \elab{eq:TDDFPT.3}{a} $z^2<0$) corresponding to real frequencies. 

If we call the zeros of \elab{eq:TDDFPT.3}{a} $\im\gO^{CM}_\gl$ we thus finally get 
\eql{eq:TDDFPT.6}
{
u_\gl\(t\)=\sum_{s=\pm}
\begin{cases}
 e^{\im s \gO^{CM}_{\gl}t}u_{\gl s} & Laplace \\
 e^{\im s \Re\[\gO^{CM}_{\gl}\]t}e ^{-\Im\[\gO^{CM}_{\gl}\]t} u_{\gl s} & Fourier
\end{cases}
}
In  \e{eq:TDDFPT.6}  the $u_{\gl s}$ constants are defined in terms of the initial position and velocity, $u_\gl^0$ and $v_\gl^0$.
\e{eq:TDDFPT.6} demonstrates that $u_\gl\(t\rar \infty\)=0$ only when the TD--DFPT master equation is solved by using the Fourier transformation. In this case 
we recover the FD--DFPT\fddfpt. \amend{\e{eq:TDDFPT.6} is schematically represented in \fig{fig:1}}.

FD--DFPT is obtained, therefore, when $u_\gl\(t\)$ and the kernel $C_\gl\(t\)$ are dumped\,(via $e^{-\eta t}$) from the beginning.
Physically this damping would describe an electronic broadening that  is zero in DFT and in DFPT. This means that
the physical solution is obtained when $\eta\rar 0$. However, in this limit the FD--DFPT produces a damped solution which corresponds to an unphysical and energy violating decay of the oscillations.
This decay is nonphysical as it makes $\average{H}$ time--dependent while the Hamiltonian, \e{eq:H.1}
is not time--dependent. More importantly, as we are in the linear regime and $f_i$ is time--independent,  decaying atomic oscillations lead to 
a decaying total
energy. This is in contrast with the fact that within the mean--field approximation the energy is conserved. 
The Laplace solution instead corresponds to persistent oscillating
solutions which represent a set of un--damped independent harmonic oscillators whose total energy is constant and conserved.

\mysec{The breakdown of the semi--classical phonon picture}
\label{sec:mbpt}
Now the natural question is what is the impact om the phonon energies of the absence, in the semi--classical TD--DFPT case,  of any energy indetermination. Are
the TD--DFPT phonon energies still reliable?

In order to answer this question we need to estimate the impact of the phonon line--width on its energy. To do this we use the MBPT approach where
the phonon frequencies are  defined as solution of the Dyson equation for the phonon Green's function,
$D_\gl\(t,t^\p\)=-\im\average{\callT\{\Delta\h{u}_\gl\(t\)\Delta\h{u}\(t^\p\)\}}$\cite{Leeuwen2013}.  The poles of the Fourier transformed $D_\gl\(-\im\go\)$
(using the convention defined by \e{eq:TDDFPT.1}) with respect to $\(t-t^\p\)$ are the MBPT phonon energies, $\gO^{QM}_\gl+\im\gc^{QM}_\gl$. Here I
use the Quantum Mechanics\,(QM) label for MBPT quantities. Those poles are solution of the fixed point equation
\eql{eq:MBPT.1}
{
 \go^2-\go^2_\gl+\go_\gl C^{e-n}_\gl-\go_\gl\Pi_\gl\(-\im \go\)=0.
}
The usual interpretation is that while $\gO^{QM}_\gl$ is the renormalized phonon energy, $\gc^{QM}_{\gl}$ defines its energy
indetermination.

In \e{eq:MBPT.1} $\Pi_\gl\(-\im\go\)$ is the Fourier transformed of the Phonon self-energy\footnote{In \e{eq:MBPT.2} I have assumed that the $g^\gl_{ij}$
screening in the MBPT case is equal to the one appearing in the TD--DFPT. This is only approximately true. For a detailed discussion see \ocite{Marini2015}.}:
\eql{eq:MBPT.2}
{
 \Pi_\gl\(-\im \go\)= \sum_{ij}g_{ji}^\gl \ti{g}_{ij}^\gl \frac{\Delta f_{ij}}{\go+\Delta\gee_{ij}+\im \xi}.
}
In \e{eq:MBPT.2} $\xi$ is a tiny positive number that appears because of the adiabatic switching on of the interaction. Thanks
to the Gell--Mann\&Low theorem~\cite{ALEXANDERL.FETTER1971} it is possible to send  $\xi\rar 0$. Thanks to this 
basic theorem of MBPT $\Pi_\gl\(-\im \go\)$ acquires a finite imaginary part and, consequently, provides the phonon with a finite energy indetermination. 

The formally analogy of \e{eq:MBPT.2} and \elab{eq:TDDFPT.3}{b} has instilled the idea that non--adiabatic effects can be described by DFPT with the same
accuracy of MBPT. From a physical point of view this would mean that a fully quantistic approach is equivalent to treat the atoms semi--classically.
Here, instead, I demonstrated that this not true and the semi--classical TD--DFPT approach has no access to the phonon energy indetermination.

\label{sec:TDDFPT_vs_MBPT}
\begin{figure}[t!]
{\centering
\includegraphics[width=\columnwidth]{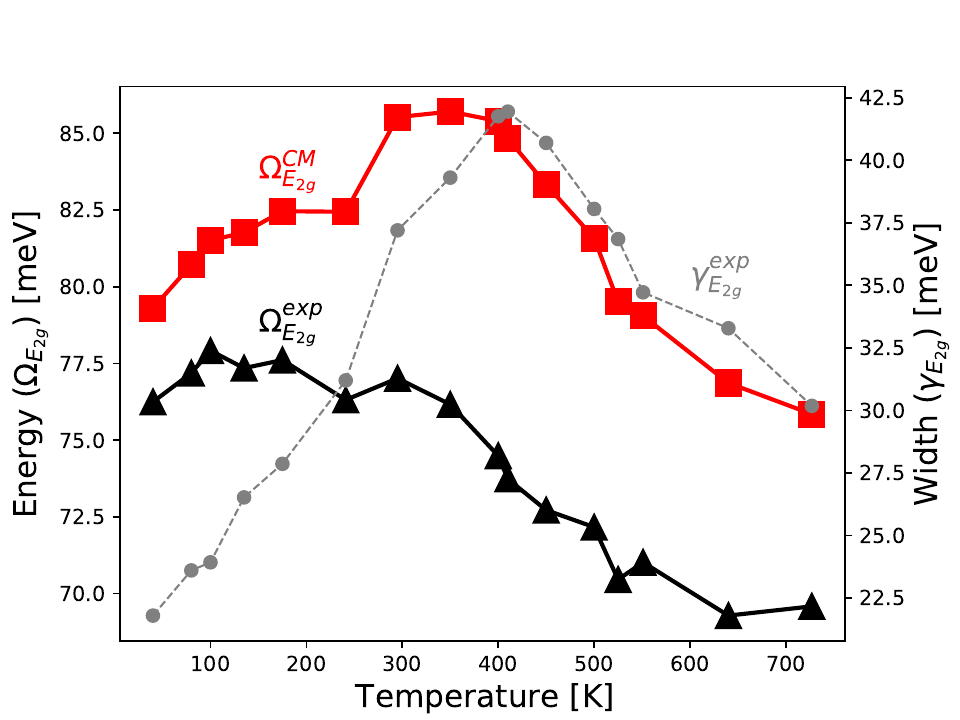}
}
\caption{MgB$_2$ experimental width ($\gc_{E_{2g}}^{exp}$, gray line, circles) and energy ($\gO_{E_{2g}}^{exp}$, black line, triangles) of the E$_{2g}$ phonon mode as a function of the
temperature\cite{Ponosov2017} compared with the semi--classical TD--DFPT phonon energies ($\gO_{E_{2g}}^{CM}$, red line, boxes) calculated by means of
\e{eq:VS.2} assuming that $\gc^{exp}_{\gl}=\gc^{QM}_{\gl}$. 
MgB$_2$ is well known to manifest strong non--adiabatic effects that result in a phonon width that, at room temperature, is almost half of the phonon energy. This very large width causes an
equally large (15\%) TD--DFPT overestimation of the phonon frequency.
}
\label{fig:2}
\end{figure}

In order to estimate the impact of the difference $\gO^{CM}_{\gl}-\gO^{QM}_{\gl}$ in realistic materials let's compare the solution of \e{eq:MBPT.1} with \elab{eq:TDDFPT.3}{a}
in MgB$_2$, a paradigmatic material characterized by very large non--adiabatic effects~\cite{Choi2002}. We can safely assume that
$\Re\[\Pi_\gl\(-\im\go\)\]\sim \callP\(C_\gl\(-\im\go\)\)$. It follows that we can rewrite the real part of  \e{eq:MBPT.1} as
\eql{eq:VS.1}
{
 \(\gO^{QM}_{\gl}\)^2+\(\gc^{QM}_{\gl}\)^2=\(\gO^{CM}_{\gl}\)^2.
}
\e{eq:VS.1} is demonstrated analytically in \smVII, and 
it represents another result of this work. It provides an easy recipe to estimate the deviation of the TD--DFPT phonon energy from the
MBPT result when the phonon acquires a finite width. First of all, we see that $\gO^{CM}_{\gl}>\gO^{QM}_{\gl}$ which means that the semi--classical phonon
frequency is always larger than the MBPT one. More importantly we can apply
\e{eq:VS.1} to any material whose experimental phonon energies and widths have been either calculated or measured experimentally.

Indeed if we suppose to use the {\em exact} MBPT self--energy, then $\gO^{QM}_{\gl}$\,($\gc^{QM}_\gl$) is, by definition, equal to the experimental
frequency\,(width). So we can use a slightly
modified version of \e{eq:VS.1}: 
\eql{eq:VS.2}
{
 \gO^{CM}_{\gl}=\sqrt{\(\gO^{exp}_{\gl}\)^2+\(\gc^{exp}_{\gl}\)^2}.
}
In the case of MgB$_2$ we can apply \e{eq:VS.2} to the $E_{2g}$ mode measured via Raman scattering by Ponoso and Streltsov in \ocite{Ponosov2017}, whose
energy\,(black line) and width\,(gray line) are reported as a function of the temperature in \fig{fig:2}.  We notice that at $T\sim 410$K 
$\gc^{exp}_{E_{2g}}$ reaches its maximum, $41.9$\,meV, that corresponds to $\sim$57\% of the frequency ($73.7$\,meV). In this case \e{eq:VS.2}
gives $\gO^{CM}_{\gl}=84.83$\,meV, which corresponds to a 15\% overestimation of the experimental value. 

In \ocite{Saitta2008} the DFPT approach has been applied to several conventional and layered metals finding deviations of the experimental phonon
frequencies from the adiabatic DFTP results ranging from $\sim 4.5$\%\,(CaC$_6$ at 300K) to $\sim 24.5$\%(KC$_8$). These deviations are of the same order of
magnitude of the $\gO^{CM}_{\gl}-\gO^{QM}_{\gl}$. This confirms that non--adiabatic effects cannot be described by TD--DFPT and impose the use of a fully
quantistic approach.

\mysec{Conclusions}
\label{sec:conclusions}
In this work I propose a time--dependent formulation of Density Functional Perturbation Theory that extends the semi--classical phonon concept to the
non--adiabatic regime.  By solving exactly the integro--differential equation which governs the atomic oscillations dynamics I have demonstrated that 
phonon energies are purely real and that a semi--classical description has no access to the phonon energy indetermination.

By comparing the TD--DFPT solution with the quasi--phonon Many--Body picture I derive a simple equation that allows to estimate, even
experimentally, the impact of the semi--classical phonon assumption. This method is applied to MgB$_2$ showing large ($15$\% of the bare phonon
frequency) overestimation of the experimental phonon energy when using TD--DFPT.
This overestimation reflects the breakdown of the semi--classical picture.  The final results is that the semi--classical DFTP phonon approach cannot describe
non--adiabatic effects. In this case a fully quantistic scheme is needed. The present results open several questions in many different fields of physics and
call for further research on the impact of non--adiabatic effects on the atomic dynamics within and beyond the harmonic approximation. 

\mysec{Acknowledgments}
A.M. would like to acknowledge F. Paleari, G. Stefanucci and E. Perfetto for helpful discussions. 
A.M. acknowledges the funding received from: 
MaX {\em MAterials design at the eXascale}, a European Centre of Excellence funded by the European Union’s program HORIZON-EUROHPCJU-2021-COE-01 (Grant No.
101093374); {\em Nanoscience Foundries and Fine Analysis -- Europe | PILOT} H2020-INFRAIA-03-2020 (Grant No. 101007417); 
{\em PRIN: Progetti di Ricerca di rilevante interesse Nazionale} Bando 2020 (Prot. 2020JZ5N9M).

\appendix
\section{The initial Hamiltonian}
\label{sec:SM_H}
The perturbed Density Functional Theory\,(DFT) Hamiltonian has been introduced by several
authors~\cite{Marini2023,Giustino2017,Leeuwen2004a,Marini2015,Stefanucci2023,Marini2018}. I take as reference eq. 24 of \ocite{Marini2023}:
\mll{eq:SM_H.1}{
\wh{H}=  \sum_{i} \gee_{i} \hat{c}^{\dagger}_{i} \hat{c}_{i}+ \wh{H}_{e-e}+\\+
\sum_{\gl}\bps{\frac{\go_{\gl}}{2}\(\h{u}^\dg_\gl\h{u}_\gl+\h{p}^\dg_\gl\h{p}_\gl\)+\\+\(\h{\callL}_\gl+ \sum_\mu \h{\callQ}_{\mu\gl} \h{\gp}_{+\mu}\) \h{\gp}_{+\gl} }
}
where $\h{\callL}_\gl$, $\h{\callQ}_{\gl\mu}$ are defined in eq. 25 of \ocite{Marini2023}.

As I am working within DFT the electron--electron interaction is replaced by the exact mean--field exchange--correlation potential, $\ul{\Delta
V}^{Hxc}$.

In the present context I assume that the equilibrium positions the atoms are oscillating around are well described
by the adiabatic DFPT, with negligible non--adiabatic corrections. Thanks to this assumption 
\eql{eq:H.2}
{
 \h{\callL}_\gl= \sum_{ij} g_{ji}^{\gl} \Delta\h{\gr}_{ij}.
}
Similarly I neglect second--order terms in \e{eq:SM_H.1} in $\h{\gp}_{s\gl}$ which induces non--harmonic corrections to the atomic equation of motion. Thanks
to this approximation
\eql{eq:H.3}
{
 \h{\callQ}_{\gl\mu}=-\frac{1}{2} C^{e-n}_{\gl\mu}.
}
I further assume $C^{e-n}_{\gl\mu}\sim C^{e-n}_{\gl\mu}\gd_{\gl\mu}$.

By using \es{eq:H.2}{eq:H.3} in \e{eq:SM_H.1} we obtain Eq.(2) of the body of this work.

The notation used in \e{eq:SM_H.1} is very general. The $i$ label can represent any one--body quantum number. If we consider, for example, the electronic momentum $\kk$ (bold symbols are vectors) \e{eq:SM_H.1}
turns into
\mll{eq:SM_H.4}{
\wh{H}=  \sum_{\kk} \gee_{\kk} \hat{c}^{\dagger}_{\kk} \hat{c}_{\kk}+ 
\sum_{\gl \qq}\bps{\frac{\go_{\qq\gl}}{2}\(\h{u}^\dg_{\qq\gl}\h{u}_{\qq\gl}+\h{p}^\dg_{\qq\gl}\h{p}_{\qq\gl}\)+\\+
\sum_{\kk} \bpr{g_{\qq}^{\gl} \h{u}_{\qq\gl}+V^{Hxc}_\qq} \h{c}^\dg_{\kk} \h{c}_{\kk-\qq} }.
}
From \e{eq:SM_H.4}, the very same procedure described in the main text leads to
\eql{eq:SM_H.5}{
 C_{\qq\gl}\(z\)=-\im g_{\qq}^\gl \ti{g}_{-\qq}^\gl \sum_\kk \frac{f_{\kk}-f_{\kk+\qq}}{z-\im\(\gee_{\kk}-\gee_{\kk+\qq}\)}.
}
\e{eq:SM_H.5} is equivalent to Eq.(145) of \ocite{Giustino2017}, for example. 
In a similar way band indexes can be added. The derivation of the main text will not depend on the specific notation used as it is completely agnostic of
the underling one--body representation.

\section{Equations of motion for the electronic and bosonic operators}
\label{sec:SM_EOMs}
The Hamiltonian defined in Eq.(2) of the main text induces a dynamics of all operators, electronic and atomic. These equations have been 
recently reviewed in \ocite{Marini2023,Marini2015,Stefanucci2023}. The second--order time derivative of the displacement is
\eql{eq:SM.eom.1}
{
\frac{\di^2}{\di t^2}\h{u}_{\gl}\(t\)=  \go_\gl\(\Pi^{st}_\gl-\go_\gl\)\h{u}_{\gl}\(t\)-\go_\gl\sum_{ij}g_{ji}^\gl \Delta\h{\gr}_{ij}\(t\),
}
while the equation of motion for the density matrix is
\mll{eq:SM.eom.2}
{
 i\frac{\di}{\di t}\h{\gr}_{ij}\(t\)=-\Delta\gee_{ij}\h{\gr}_{ij}\(t\)\\+\sum_{\gl}
\[\ul{\h{\gr}}\(t\),\ul{g}^\gl\h{u}_\gl\(t\)\]_{ij}+\[\ul{\h{\gr}}\(t\),\Delta\ul{V}^{Hxc}\[\gr\(t\)\]\]_{ij}
}
\es{eq:SM.eom.1}{eq:SM.eom.2} completely solve the Many--Body problem and produce the equations governing the classical atomic motion as an
approximation as explained in main text.

\section{Electron--phonon interaction dynamical dressing}
\label{sec:g_dressing}
The role of $V^{Hxc}$ in \e{eq:SM.eom.2} is to  screen the electron--phonon interaction $g^{\gl}_{ji}$. In order to see it we observe
that~\cite{978-3-642-23518-4}
\mll{eq:SM_g_dress.1}
{
 \ul{\Delta V}^{Hxc}\[\gr\]\(t\)= 
 \int \di \tau\, t^\p \ul{\ul{f}}^{Hxc}\(t-\tau\) \gd \ul{\gr}\(\tau\)=\\
 \sum_\gl \int \di \tau\, t^\p \ul{\ul{f}}^{Hxc}\(t-\tau\) \ul{\ul{\chi}}\(\tau-t^\p\) \ul{g}^\gl u_\gl\(t^\p\),
}
with $\ul{\ul{\chi}}$ the tensorial reducible response function and $\ul{\ul{f}}^{Hxc}$ the Hartree plus exchange--correlation kernel, defined as
$\frac{\gd \ul{V}^{Hxc}\[\gr\]\(t\)}{\gd \ul{\gr}\(t^\p\)}$. The single underlined
quantities are matrices\,($\ul{M}_{ij}$) while the doubly underlined are tensors\,($\ul{\ul{M}}_{\sst{ij}{kl}}$) in the electron--hole pairs space. If we now
plug \e{eq:SM_g_dress.1} in \e{eq:SM.eom.2} and take the average of both terms we get
\mll{eq:SM_g_dress.2}
{
 \ul{g}^\gl u_\gl\(t\)+\ul{\Delta V}^{Hxc}\[\gr\(t\)\]=\sum_\mu \int \di t^\p 
 \bps{\ul{\ul{\gd}}\gd_{\gl\mu}\gd\(t-t^\p\)+\\+\int d\tau \ul{\ul{f}}^{Hxc}\(t-\tau\) \ul{\ul{\chi}}\(\tau-t^\p\)} \ul{g}^{\mu}
 u_{\mu}\(t^\p\)\sim\\
 \int\di t^\p \ul{\ul{\varepsilon}}^{-1}\(t-t^\p\) \ul{g}^{\gl}u_\gl\(t^\p\).
}
In \e{eq:SM_g_dress.2} 
it appears the TD--DFT inverse test--electron~\cite{Marini2015} dielectric matrix, $\ul{\ul{\varepsilon}}^{-1}\(t-t^\p\)$, and I have neglected the  $\gl\neq\mu$ terms.

If we now use \e{eq:SM_g_dress.2} in \e{eq:SM.eom.2} we get 
\eql{eq:SM_g_dress.3}{
 \Delta \ul{\gr}\(t\)=\int \di t^\p \ul{\ul{\chi}}\(t-t^\p\)\ul{g}^{\gl} u_\gl\(t^\p\),
}
that corresponds to the TD--DFT Kubo equation. If we now use the K\"all\'en--Lehmann spectral representation of $\ul{\ul{\chi}}\(t-t^\p\)$~\cite{ALEXANDERL.FETTER1971} we can expand
\e{eq:SM_g_dress.3} over the poles ($E_I$) and residuals ($\ul{\ul{\chi}}^I$) of $\ul{\ul{\chi}}$:
\eql{eq:SM_g_dress.4}{
 \gr_{ij}\(t\)=\sum_I \[\ul{\ul{\chi}}^I\ul{g}^{\gl}\]_{ij} \int \di t^\p e^{\im E_I\(t-t^\p\)} u_\gl\(t^\p\),
}
From \e{eq:SM_g_dress.4} it follows that the effect of the time--dependent exchange--correlation DFT potential is to replace the independent particle
energies with the poles of the full TD--DFT response function. Again, \e{eq:SM_g_dress.4} does not lead to any change in the main derivations of the work
where, for simplicity, I use the static screening approximation and write
\eql{eq:SM_g_dress.5}{
 \gr_{ij}\(t\)\approx \ti{g}^{\gl}_{ij} \int \di t^\p e^{\im \Delta \gee_{ij}\(t-t^\p\)}u_\gl\(t^\p\),
}
with $\ul{\ti{g}}^{\gl}\sim \ul{\ul{\varepsilon}}^{-1}\(\go=0\) \ul{g}^{\gl}$.

\section{On the Laplace and Fourier transformations of the differential operator}
\label{sec:laplace_vs_fourier}
The Laplace and Fourier transformations are, apparently, very similar:
\eqgl{eq:SM.damp.A.1}
{
 g^{f}\(\go\)=\int_0^\infty e^{\im \go t} g\(t\)\di t,\\
 g^{l}\(\go\)=\int_0^\infty e^{-\go t} g\(t\)\di t.
}
In \e{eq:SM.damp.A.1} I have assumed $g\(t\)=0$ when $t<0$, which is the case of a classical pendulous displaced from the equilibrium position at $t=0$.

The Fourier transformation is not listed among the solvers of the integro--differential Volterra equation\cite{Wazwaz2011,boas_mathematical_2006} and the reason is simple. If we apply
\elab{eq:SM.damp.A.1}{a} to the differential operator ($\frac{\di^2}{\di t^2}$) we get
\mll{eq:SM.damp.A.2}
{
 \int_0^\infty e^{\im \go t} \frac{\di^2 u_\gl\(t\)}{\di t^2}\di t=\left. e^{\im \go t} \frac{\di u_\gl\(t\)}{\di t}\right|_{0}^{\infty}\\
 -\im\go\[ \left. e^{\im \go t} u_\gl\(t\)\right|_{0}^{\infty}-\im\go u_\gl^f\(\go\)\],
}
with $u_\gl^f\(\go\)$ the Fourier transformation of $u\(t\)$.
From \e{eq:SM.damp.A.2} it is evident that in order for the differential operator to be Fourier transformable we do need to impose the condition
$\evalat{\frac{\di u_\gl\(t\)}{\di t}}{t=\infty}=u_\gl\(t=\infty\)=0$. Therefore the Fourier transformation can be used if and only if 
it is assumed from the beginning that the solution will decay in time.

This a peculiar property of the Volterra equation: if we start from the sub--space of function that decay for $t\rar\infty$ the solution will decay as well even
if at the end of the derivation we extend the initial sub--space of functions to the entire space. 

\section{Regularization of the Laplace transformation of the time--dependent DFPT kernel}
\label{sec:regularization}
We now notice that $C_\gl\(\im\Delta\gee_{km}\)=\infty$ for any $\Delta\gee_{km}$. This means that $u_\gl\(z\)$ 
 is ill defined when $z$ approaches the electron--hole energies. However let's rewrite $C_\gl\(\go\)$ as
\eql{eq:LF.1}
{
 C_\gl\(\go\)=\(-\im\)\frac{ \sum_{J}R^\gl_J \prod_{I\neq J}\(\go-\im\Delta\gee_{I}\)}{\prod_{I}\(\go-\im\Delta\gee_{I}\)},
}
with $I=\(i,j\)$ and $R_I=g_{ji}^\gl \ti{g}_{ij}^\gl \Delta f_{ij}$. If we now use \e{eq:LF.1} to rewrite $u_\gl\(z\)$ we get
\begin{widetext}
\mll{eq:LF.2}
{
 \frac{u_\gl\(z\)}{\(z u_\gl^0 +v_\gl^0\)} = \[z^2-\go_\gl\(C_\gl^{e-n}-\go_\gl\)+\go_\gl C_\gl\(z\)\]^{-1}=\\
\frac
{\prod_{I}\(\go-\im\Delta\gee_{I}\)}
{\[z^2-\go_\gl\(C_\gl^{e-n}-\go_\gl\)\]\prod_{I}\(\go-\im\Delta\gee_{I}\)-\im \go_\gl  \sum_{J}R^\gl_J \prod_{I\neq J}\(\go-\im\Delta\gee_{I}\)}.
}
\end{widetext}
From \e{eq:LF.2} it follows that, if we use $\Delta\gee_{km}\rar \Delta\gee_K$ we get
\eql{eq:LF.3}
{
u_\gl\(\im\Delta\gee_{K}\)=\frac{\im}{\go_\gl}\frac{0}{R^\gl_K \prod_{I\neq K}\im\(\Delta\gee_K-\Delta\gee_{I}\)}=0.
}
This means that the points $z=\im\Delta\gee_{K}$ can be safely excluded
from the complex plane integral as they do not give any contribution. This implies that we 
can replace $C_\gl\(z\)$ with its {\em Cauchy principal value}, $C_\gl\(z\)\rar \callP\[C_\gl\(z\)\]$ which leads to a well
defined Bromwhich integral. 

\section{Solution of the TD--DFPT and MBPT fixed--point equations}
\label{sec:QPH}
Let's start from the TD--DFPT and MBPT equations for the phonon energy and width:
\eqgl{eq:SM_QPH.1}
{
 z^2   +\go^2_\gl+\go_\gl \gD C_\gl\(z\)=0,\\
 \go^2-\go^2_\gl+\go_\gl C^{e-n}_\gl-\go_\gl\Pi_\gl\(-\im \go\)=0.
}
We now rotate the Laplace equation to the imaginary axis $z\rar -\im \go$ so that the first equation becomes:
\eql{eq:SM_QPH.2}
{
 -\go^2 +\go^2_\gl+\go_\gl \gD C_\gl\(-\im \go\)=0.
}
I now introduce a quasi--phonon form of the energy dependence of $C_\gl$ and $\Pi_\gl-C^{e-n}_\gl$:
\eql{eq:SM_QPH.3}
{
 \Pi_\gl\(-\im \go\)-\go_\gl C^{e-n}_\gl\sim \ga_\gl \frac{\go^2}{\go_\gl}+\im \gb_\gl \go,
}
with $\ga_\gl$ and $\gb_\gl$ to constants that can be easily defined by comparing \e{eq:SM_QPH.3} with the analytic expression of $\Pi_\gl$.

We now assume $\gD C_\gl\(-\im \go\) \approx \callP \[ \Pi_\gl\(-\im \go\)-C^{e-n}_\gl\]$. This is a very good approximation as the phonon self--energy
and the TD--DFPT kernel have a very similar analytic form. If we define $\gO_\gl^{CM}$ the solution of \e{eq:SM_QPH.2}, from \e{eq:SM_QPH.3}  it follows that
\eql{eq:SM_QPH.4}
{
 \(\gO_\gl^{CM}\)^2=\frac{\go_\gl^2}{1-\ga_\gl}.
}
Similarly if we define $\gO^{QM}_\gl+\im\gc^{QM}_\gl$ the solution of \elab{eq:SM_QPH.1}{b} we get
\eqgl{eq:SM_QPH.5}
{
 \(\gO^{QM}_\gl\)^2-\(\gc^{QM}_\gl\)^2=\(\gO_\gl^{CM}\)^2-\frac{\gb_\gl \gc^{QM}_\gl \go_\gl}{1-\ga_\gl},\\
 \gc^{QM}_\gl=\frac{\gb_\gl\go_\gl}{2\(1-\ga_\gl\)}.
}
If we now replace \elab{eq:SM_QPH.5}{b} in  \elab{eq:SM_QPH.5}{a} we get
\eql{eq:SM_QPH.6}
{
  \(\gO^{QM}_\gl\)^2+\(\gc^{QM}_\gl\)^2=\(\gO_\gl^{CM}\)^2.
}

\bibliography{paper}

\end{document}